\documentclass[runningheads]{llncs}
\usepackage{amsmath}
\usepackage{amssymb}
\usepackage{graphicx}
\usepackage{booktabs}
\usepackage{array}
\usepackage[all]{xy}
\usepackage{tikz}
\usepackage{verbatim}
\usepackage[left=3cm,right=3cm,top=3cm,bottom=2.5cm]{geometry}
\usepackage{hyperref}
\usepackage{caption}
\usepackage{subcaption}
\usepackage{psfrag}
\usepackage{actuarialangle}

\begin{document}
\title{Efficient Novelty Detection Methods for Early Warning of Potential Fatal Diseases\thanks{Supported by the African Institute for Mathematical Sciences - Rwanda.}}

\titlerunning{Early Warning of Potential Fatal Diseases}
%
\author{Sèdjro Salomon Hotegni\inst{1}
\and
Ernest Fokoué\inst{2}
}

\authorrunning{S. Hotegni \and E. Fokoue}
%
\institute{African Institute for Mathematical Sciences, Rwanda 
\and	
\email{salomon.hotegni@aims.ac.rw}\\
\and
Rochester Institute of Technology, United States\\
\email{epfeqa@rit.edu}}
\maketitle              
\begin{abstract}
Fatal diseases, as Critical Health Episodes (CHEs), represent real dangers for patients hospitalized in Intensive Care Units. These episodes can lead to irreversible organ damage and death. Nevertheless, diagnosing them in time would greatly reduce their inconvenience. This study therefore focused on building a highly effective early warning system for CHEs such as Acute Hypotensive Episodes and Tachycardia Episodes. To facilitate the precocity of the prediction, a gap of one hour was considered between the observation periods (Observation Windows) and the periods during which a critical event can occur (Target Windows). The MIMIC II dataset was used to evaluate the performance of the proposed system. This system first includes extracting additional features using three different modes. Then, the feature selection process allowing the selection of the most relevant features was performed using the Mutual Information Gain feature importance. Finally, the high-performance predictive model LightGBM was used to perform episode classification. This approach called MIG-LightGBM was evaluated using five different metrics: Event Recall (ER), Reduced Precision (RP), average Anticipation Time (aveAT), average False Alarms (aveFA), and Event F1-score (EF1-score). A method is therefore considered highly efficient for the early prediction of CHEs if it exhibits not only a large aveAT but also a large EF1-score and a low aveFA. Compared to systems using Extreme Gradient Boosting, Support Vector Classification or Naive Bayes as a predictive model, the proposed system was found to be highly dominant. It also confirmed its superiority over the Layered Learning approach.

\end{abstract}
\section{INTRODUCTION}
In healthcare, the purpose of an early warning system for a Critical Health Event is to provide health specialists and the public with as much notice as possible of the event's probability of occurrence, thereby extending the range of viable responses. It’s of great importance to the reduction of mortality in healthcare.
In Rwanda, for instance, the Intensive Care Units (ICUs) mortality rate is around $43.8\%$, which is exceedingly high \cite{ref1}. The ICUs are special units where patients with life-threatening health issues are monitored and treated, and support units are provided when necessary to maintain vital functions. Therefore, they represent a direct target for Critical Health Episodes. Adverse events (CHEs) that frequently occur in the ICUs include the deterioration of the patient's situation due to a failure to rescue or act in a timely manner. Acute Hypotensive Episodes (AHE) and Tachycardia Episodes (TE) are two of the most dangerous Critical Health Episodes in the Intensive Care Units.

An Acute Hypotensive Episode is defined as any interval of $30$ minutes or longer during which at minimum $90\%$ of the Mean Arterial Pressure (MAP) values were at or below $60$ mmHg \cite{ref2}.
Tachycardia Episodes, on the one hand, can be defined as any interval of $30$ minutes or more during which at least $90\%$ of heart rate measurements are over 100 bpm \cite{ref4}.

The specific objective of this study is to provide clinicians with at least one-hour warning of an imminent AHE or TE by providing a warning system that is extremely fast, uses a wide range of information from currently accessible patient monitoring data, and surpasses existing models of AHE and TE early detection.
\section{METHODOLOGY}
In general, existing warning systems for AHE and TE early prediction typically have limited performance and high computational costs in real-time alerting when dealing with large amounts of data. So, this work presents MIG-LightGBM, a highly efficient warning system based on a feature selection process with Mutual Information Gain (MIG) and an episode classification approach with the predictive model Light Gradient Boosting Machine (LightGBM).(Figure \ref{fig3})
\begin{figure}[!ht]
	\centering
	\includegraphics[scale=0.4]{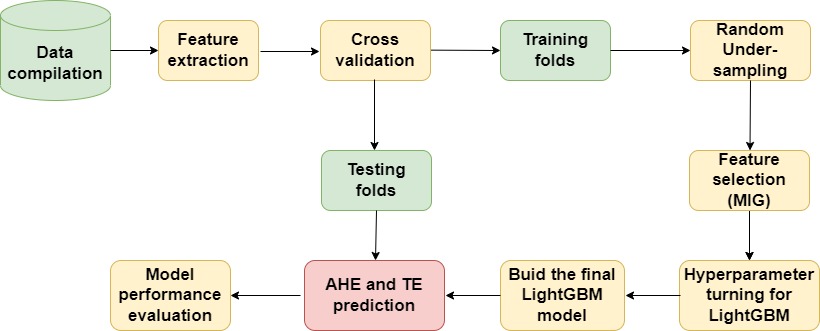}
	\caption{Proposed early AHE and TE prediction system (MIG-LightGBM)}
	\label{fig3}
\end{figure}

\subsection{Data compilation}
\begin{figure}[!ht]
	\centering
	\includegraphics[scale=0.4]{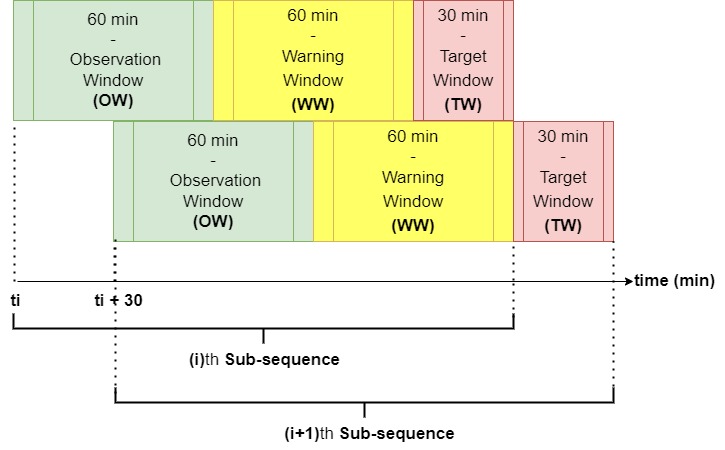}
	\caption{Example of two consecutive sub-sequences}
	\label{fig2}
\end{figure}

The data used in this research is a subset of the Multi-parameter Intelligent Monitoring for Critical Care (MIMIC) II database \cite{ref6}. It contains minute-by-minute time series of Heart Rate (HR), Systolic Blood Pressure (SBP), Diastolic Blood Pressure (DBP), and Mean Arterial blood Pressure (MAP) arranged into records, each of which corresponds to
an adult patient’s ICU stay.\\
The dataset was traversed by sub-sequences, each one being subdivided into three windows (Figure \ref{fig2}).\\
\textit{Target Window (TW):} It is the period during which a Critical Health Event can occur.\\
\textit{Observation Window (OW):} OW is the Observation period containing data to be used to predict what will happen in the Target Window.\\
\textit{Warning Window (WW):} It is the gap between the Observation Window and the Target Window.

Following Lee and Mark\cite{ref7}'s data compilation procedure, the values of HR (in bpm) and MAP (in mmHg) in each $30$ minutes target window must be between $10$ and $200$; otherwise, the corresponding sub-sequence was eliminated from this research.
In addition, for all the four signals to be regarded valid and therefore taken into consideration for this study, their values must be at least $95\%$ between $10$ and $200$ in the $60$ minutes Observation Windows. If not, the corresponding sub-sequence is also eliminated from the analysis.
Finally, $86$ different patient records were used for the analysis.

Using the target window of each episode, two feature targets $\textbf{y}_1$ and $\textbf{y}_2$ were added to the compiled time series data of each patient.
\newline
The variable $\textbf{y}_1$ is binary and indicates whether a sub-sequence is a Hypotensive Episode or not. 
Similarly, $\textbf{y}_2$ is binary and indicates whether a sub-sequence is a Tachycardia Episode or not.
So, for both AHE and TE, the early prediction task was seen as a classification task.

A 5-fold Cross-Validation (CV) was performed on the dataset, considering patients instead of observations (each fold contains patients whose data was used for analysis, some for training and the rest for testing).
To permit the classifier to properly learn the unbalanced data, each data in the training folds was under-sampled 10 times without replacement.The Test data, on the other hand, were left imbalanced.

\subsection{Feature extraction}
In this work, the feature extraction process was carried out according to the techniques used by Tsur et al.\cite{ref7} and Lee and Mark\cite{ref9}. Using the four existing time series features, two knowledge-based features such as Pulse Pressure (PP) and Cardiac Output (CO) were derived. Then, for each of these 6 prior time series features, 10 statistical features were extracted based on the Observation Windows. The cross-correlation technique was used to compare the 6 time series features two-by-two and objectively determines how well they match each other ($C_6^2 = 15$ additional features). In addition, the Meyer wavelet was used to perform a 5-level discrete wavelet decomposition of each of the 6 signals to extract relative energies in distinct wavelength bands ($5 × 6 = 30$ new features).
The total number of features considered for the analysis was therefore $111$.

\subsection{Feature selection}
Apart from the feature importance ranking
provided by the proposed predictive model, the Mutual Information Gain approach was also used for the feature selection process.
\paragraph{Mutual Information Gain (MIG):}
It is a measure of the "mutual dependency" of two random variables \cite{ref10}. MIG creates a quantifiable link between a feature and the target. Using MIG as a feature selector has two advantages: It is model-neutral, which means it can be applied to a wide range of machine learning models; and it is also fast.

Let \textbf{$\text{x}_j$} be a feature and \textbf{$\text{y}$} the target variable. The Mutual Information Gain for the two discrete random variables $\textbf{$\text{x}_j$}$ and \textbf{$\text{y}$} is given by
\begin{equation}
	I(\textbf{$\text{x}_j$};\textbf{$\text{y}$}) = I(\textbf{$\text{y}$};\textbf{$\text{x}_j$}) = \sum_{y_i\in \textbf{$\text{y}$}}\sum_{x_{ji}\in \textbf{$\text{x}_j$}}p_{xy}(x_{ji},y_i).log\bigg( \dfrac{p_{xy}(x_{ji},y_i)}{p_x(x_{ji})p_y(y_i)}\bigg)
\end{equation}

where $p_x$ and $p_y$ are the marginal probability and $p_{xy}$ is the joint probability.

Given a set of features $\mathcal{X}_{\mathcal{S}} = \{\textbf{$\text{x}$}_{1},...,\textbf{$\text{x}$}_{n}\}$ and a single feature $\textbf{y}$, the \textit{Joint Mutual Information} between them is given by
\begin{equation}
	I(\mathcal{X}_{\mathcal{S}};\textbf{y}) = I(\textbf{$\text{x}$}_{i_1},...,\textbf{$\text{x}$}_{i_{k}};\textbf{y}) = \sum_{j}I(\textbf{$\text{x}$}_{j};\textbf{y}|\textbf{$\text{x}$}_{j-1}, \textbf{$\text{x}$}_{j-2},...,\textbf{$\text{x}$}_{1}).
\end{equation}

Let $|\mathcal{S}|=k.$ be the number of features to be selected.
\newline
The feature selection process should  identify a subset of features $\mathcal{X}_{\hat{\mathcal{S}}} = \{\textbf{$\text{x}$}_{i_1},...,\textbf{$\text{x}$}_{i_{k}}\}$, which maximizes the Joint Mutual Information $I(\mathcal{X}_{\mathcal{S}};\textbf{y}))$ between the class label $\textbf{y}$ and all possible feature subsets $\mathcal{X}_   \mathcal{S}$ of size k
\begin{equation}
\{i_1,...,i_{k}\} = \hat{\mathcal{S}} = arg\underset{\mathcal{S}}{max}\{ I(\mathcal{X}_{\mathcal{S}};\textbf{y})\}
\end{equation}

For this, the \textit{Greedy forward step-wise selection} method was used.
According to their Mutual Information with respect to the target $\textbf{y}$, the features are ranked. The top feature is then selected.
Let $\mathcal{X}_{\mathcal{S}^{t-1}}=\{\textbf{$\text{x}$}_{i_1},...,\textbf{$\text{x}$}_{i_{t-1}}\}$ denote the set of features that were chosen at step $t-1$. The next feature $\textbf{x}_{i_t}$ is chosen in such a way that the greatest improvement in Joint Mutual Information is achieved by using $\mathcal{X}_{\mathcal{S}^t}$. So,
\begin{equation}
	i_t = arg\underset{i\notin \mathcal{S}^{t-1}}{max}I(\mathcal{X}_{\mathcal{S}^{t-1}\cup i};\textbf{y}) = arg\underset{i\notin \mathcal{S}^{t-1}}{max}I(\mathcal{X}_{\mathcal{S}^{t-1}}, \textbf{$\text{x}_i$};\textbf{y})
\end{equation}
The optimization problem is simplified for a given $\alpha$ and $\beta$ in $[0;1]$ to:

\begin{equation}
	i_t = arg\underset{i\notin \mathcal{S}^{t-1}}{max}\bigg(
	\underbrace{I(\textbf{$\text{x}_i$};\textbf{y})}_\text{relevancy}-\bigg[ \underbrace{\alpha\sum_{j\in \mathcal{S}^{t-1}}I(\textbf{$\text{x}_j$};\textbf{$\text{x}_i$})-\beta\sum_{j\in \mathcal{S}^{t-1}}I(\textbf{$\text{x}_j$};\textbf{$\text{x}_i$}|\textbf{y})}_\text{redundancy}\bigg]\bigg)
\end{equation}
Generally, a value of $\alpha$ near-zero is related to the hypothesis that all features are class-conditionally independent of each other and a value of $\beta$ near-zero is related to the hypothesis that all features are independent of one another.

\subsection{Predictive model: Light Gradient Boosting Machine}
Light Gradient Boosting Machine (LightGBM) is a sophisticated \textit{Gradient-Boosted Decision Tree (GBDT)} method that is designed to identify the optimal feature splitting points while also reducing the quantity of samples and features.\\
\textit{Gradient-based One Side Sampling (GOSS)} and \textit{leaf-wise growth} are its two key benefits.

\paragraph{Gradient-based One Side Sampling (GOSS)}
The key insight of GOSS is that data instances with stronger gradients play
larger roles in information gain computation. So, it preserves them and randomly picks data with small gradients when determining the optimal split. To do so:
\begin{itemize}
	\item First, GOSS ranks the training instances based on the absolute values of their gradients. 
	\item Then, it selects the top $a\%$ of the total instances with the largest gradient. (a subset $A$ is thus obtained)
	\item Random sampling $b\%$ of instances from the remaining $(1-a)\%$.  (another subset $B$ is obtained)
	\item The gradients of the $b\%$ of instances are multiplied by $\dfrac{(1-a)}{b}$, which amplifies the lower gradients.
\end{itemize}

Finally, the instances are split based on the projected variance gain $\hat{V_j}(d)$ over the subset of selected instances:
\begin{align*}
	\hat{V_j}(d) &= \dfrac{1}{n}\bigg(\dfrac{\big(\sum_{\textbf{$\text{x}_i$}\in A_l}k_i+\dfrac{1-a}{b}\sum_{\textbf{$\text{x}_i$}\in B_l}k_i\big)^2}{n_l^j(d)} + \dfrac{\big(\sum_{\textbf{$\text{x}_i$}\in A_r}k_i+\dfrac{1-a}{b}\sum_{\textbf{$\text{x}_i$}\in B_r}k_i\big)^2}{n_r^j(d)}\bigg)
\end{align*}

where,\\
$A_l=\{\textbf{$\text{x}_i$}\in A|x_{ij}\le d\}$, $B_l=\{\textbf{$\text{x}_i$}\in B|x_{ij}\le d\}$,
$A_r=\{\textbf{$\text{x}_i$}\in A|x_{ij}> d\}$, $B_r=\{\textbf{$\text{x}_i$}\in B|x_{ij}> d\}$,
\newline
$n_l^j(d)=|\{\textbf{$\text{x}_i$}\in A\cup B|x_{ij}\le d\}|$,
$n_r^j(d)=|\{\textbf{$\text{x}_i$}\in A\cup B|x_{ij}> d\}|$,
and $\{k_1,...,k_n\}$ are the negative gradients of the loss function with respect to the output of the model.

On the subset $A\cup B$, for each feature $j$, GOSS selects $d_j^*=\underset{d}{argmax}\hat{V_j}(d)$. The data is then split based on the feature $j^* = \underset{j}{argmax}\hat{V_j}(d_j^*)$ at the point $d_j^*$.

\paragraph{leaf-wise tree growth:} In LightGBM, the \textit{leaf-wise tree growth} technique chooses the leaf that minimizes loss the most and splits only that leaf, ignoring the rest of the leaves at the same level.
As a result, the tree becomes asymmetrical, and additional splitting may occur only on one side of the tree. (Figure \ref{fig3})

\begin{figure}[!ht]
\centering
\includegraphics[scale=1]{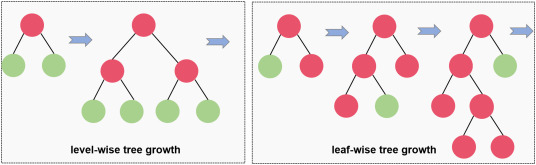}
\caption{Level-wise tree growth VS leaf-wise tree growth}
\label{fig3}
\end{figure}

There are another two reasons why LightGBM is very fast: \textit{Histogram-based splitting} and \textit{Exclusive Feature Bundling (EFB)}.

\paragraph{Histogram-based splitting:}
It separates the data into a given number of bins of uniform length, and then iterates across these bins to determine the best split point.
So, the complexity of the algorithm goes from $\mathcal{O}(data \times features)$ to $\mathcal{O}(bins \times features)$, with $bins \ll data$.
	
\paragraph{Exclusive Feature Bundling:}
This technique identifies and combines the mutually exclusive features (features that never take zero values simultaneously) into a single feature to decrease the complexity to $\mathcal{O}(bins \times bundle)$, with $bundle \ll features$.

\paragraph{}
In order to conduct a comparative study on LightGBM and other predictive models, the \textit{Extreme Gradient Boosting (XBoost)}, \textit{Naive Bayes (NB)}, and \textit{Support Vector Classification (SVC)} have also been used. In addition, the \textit{Random Search} hyperparameter tuning method was performed to choose the best parameters of the models.

\subsection{Evaluation Metrics}
To evaluate the performance of the predictive models, following Cerqueira et al.\cite{ref4} four metrics were assessed first: Event Recall (ER), Reduced Precision (RP); the average number of False Alarms aveFA, and the average Anticipation Time aveAT (the average time left in advance by a model to correctly predict a Positive Case).

\paragraph{Event Recall (ER): }
\label{ER}
The Recall of a model assesses its ability to recognize positive cases. The more positive cases identified, the larger the Recall. In this study, a case corresponds to a whole subsequence instead of an individual observation. Therefore, the classical Recall metric can be misleading. An example to better understand is the case where the model identifies several individual positive observation in the same subsequence. The classical Recall metric would consider these predictions as different predictions while they refer to the same event. To avoid this, the Event Recall measure considers only the first predicted positive observation of a subsequence as sufficient to classify this subsequence as a positive case. During the computation of the Recall, this approach then ignores any other positive alarm following the first one, in the same subsequence.
The following equation calculates the $ER$ for a model $m$:
\begin{equation}
	ER_m = \dfrac{\hat{T}_m}{T},
\end{equation}
where $\hat{T}_m$ denotes the total number of the  correctly predicted events, and $T$ is the total number of the main events in the test data. 

\paragraph{Reduced Precision (RP): }
The Precision of a model is described as the proportion of properly recognized positive cases (True Positive) to the overall number of positively classified cases (either True Positive or False Positive). Because of the scenario stated in \nameref{ER}, the classical Precision metric can be misleading. So, RP also considers the number $\hat{T}_m$ of the  correctly predicted events as the number of properly recognized positive cases. Similarly, consider the scenario where several false alarms appear in the same subsequence. RP avoids making a mistake by considering the first false alarm as sufficient to count the corresponding subsequence as a False Positive case. So, the RP metric changes the number of False Positives with the number of Discounted False Positives: the number of subsequences related to the False Positive alarms.
The following equation gives the \textit{RP} for a model \textit{m}:
\begin{equation}
	RP_m = \dfrac{\hat{T}_m}{\hat{T}_m+DFP_m}
\end{equation}
where $DFP_m$ is  the number of Discounted False Positives.

\paragraph{} In addition, we added another metric to easily identify the best model.

\paragraph{Event F1-score (EF1-score): }
The EF1-score is defined here as the harmonic mean of the Reduced Precision and the Event Recall. It uses the following formula to merge RP and ER into a single value:

\begin{equation}
	\text{EF1-score} = 2\times\dfrac{ER\times RP}{ER+RP}.
\end{equation}

The goal is to maximize both Event Recall and Reduced Precision such that given pairs of their values for various models, one could compare and determine which is the best.

In this study, a model is therefore considered to be a highly efficient algorithm for the early prediction of Acute Hypotensive Episodes and Tachycardia Episodes, if it presents a high Event F1-score (EF1-score) with a large average Anticipation Time (aveAT), and a low average False Alarms (aveFA).
\section{Results}
There are three sorts of outcomes for both early prediction of AHE and TE, expressed in terms of the mean and the standard deviation of each Evaluation Metric. Note that the average Anticipation Time (aveAT) was computed here in minutes for each model.

\subsection{AHE Early Prediction}
Considering the patient data in each of the 5 folds obtained after a 5-fold Cross-Validation, the statistics presented in the Tables \ref{TrainDataAHE} and \ref{TestDataAHE} were obtained for the Train set and the Test set.

\begin{table}[!h]
	\begin{subtable}[b]{.5\linewidth}
	\centering
	\caption{Train data for AHE}
	\label{TrainDataAHE}
	\begin{tabular}{llllll}
		\toprule
		{} & Fold 1 & Fold 2 & Fold 3 & Fold 4 & Fold 5 \\
		AHE                 &        &        &        &        &        \\
		\midrule
		OW &   2718 &   2337 &   2497 &   2757 &   2535 \\
		Main Events         &    286 &    188 &    206 &    255 &    241 \\
		\bottomrule
	\end{tabular}
	\end{subtable}
    \hfill
	\begin{subtable}[b]{.5\linewidth}
		\centering
		\caption{Test data for AHE}
		\label{TestDataAHE}
\begin{tabular}{llllll}
	\toprule
	{} & Fold 1 & Fold 2 & Fold 3 & Fold 4 & Fold 5 \\
	AHE                 &        &        &        &        &        \\
	\midrule
	OW &    903 &   1277 &    916 &    603 &   1153 \\
	Main Events         &     13 &    122 &     94 &     43 &     63 \\
	\bottomrule
\end{tabular}
	\end{subtable}
\end{table}

So each fold of the Training set contains at least $2,337$ Observation Windows ($2,337\ \text{hours}$) with at least $188$ of them related to the Acute Hypotensive Episodes. In addition, each fold of the Test set contains at least $603$ Observation Windows ($603\ \text{hours}$) with at least $13$ of them related to the Acute Hypotensive Episodes.

\subsubsection{Full Dataset:}
After evaluating the performance of the predictive models XGBoost, LightGBM, SVC and Naive Bayes, the average results obtained across the 5 folds are presented in the Table \ref{AHE_res_full} considering the full data.

\begin{table}[!h]
	\centering
	\caption{AHE: Average Results on the Full Data (mean $\pm$ std)}
	\label{AHE_res_full}
	\begin{tabular}{llllll}
		\toprule
		{} &             ER &             RP &           aveFA &           aveAT &      EF1-score \\
		Models   &                &                &                 &                 &                \\
		\midrule
		XGBoost  &   0.687$\pm $0.12 &  0.364$\pm $0.177 &   2.585$\pm $0.836 &   46.781$\pm $7.69 &  0.462$\pm $0.184 \\
		LightGBM &  0.695$\pm $0.122 &    \textbf{0.37$\pm $0.18} &   \textbf{2.376$\pm $0.677} &  46.813$\pm $6.426 &  \textbf{0.469$\pm $0.186} \\
		SVC      &  0.648$\pm $0.245 &  0.067$\pm $0.035 &  25.18$\pm $11.531 &  45.939$\pm $7.896 &   0.12$\pm $0.063 \\
		NB       &   \textbf{0.946$\pm $0.05} &  0.116$\pm $0.047 &  21.256$\pm $5.235 &  \textbf{56.595$\pm $1.611} &  0.204$\pm $0.077 \\
		\bottomrule
	\end{tabular}
\end{table}

LightGBM has the highest Reduced Precision $(0.38\pm 0.185)$ and the lowest average False Alarms $(2.419\pm 0.683)$. Although it ranks second behind Naive Bayes for the Event Recall and the average Anticipation Time, it holds the highest Event F1-score $(0.478\pm 0.189)$.
The XGBoost model, on the other hand, is behind LightGBM for each metric.

\subsubsection{Feature Selection with LightGBM:}
Using the \textit{Split importance} method for the LightGBM feature importance ranking, features with non-zero importance were selected. The average results obtained across the 5 folds are then presented in the Table \ref{AHE_res_Light}.

\begin{table}[!h]
	\centering
	\caption{AHE: Average Results After LightGBM Feature Selection (mean $\pm$ std)}
	\label{AHE_res_Light}
	\begin{tabular}{llllll}
		\toprule
		{} &             ER &             RP &           aveFA &           aveAT &      EF1-score \\
		Models   &                &                &                 &                 &                \\
		\midrule
		XGBoost  &  0.686$\pm $0.124 &  0.359$\pm $0.178 &   2.684$\pm $0.685 &   47.374$\pm $6.63 &  0.458$\pm $0.189 \\
		LightGBM &  0.708$\pm $0.095 &   \textbf{0.38$\pm $0.185} &   \textbf{2.419$\pm $0.683} &  48.668$\pm $2.932 &  \textbf{0.478$\pm $0.189} \\
		SVC      &   0.834$\pm $0.04 &  0.079$\pm $0.033 &  34.063$\pm $6.254 &  55.663$\pm $3.745 &  0.143$\pm $0.057 \\
		NB       &  \textbf{0.959$\pm $0.031} &   0.14$\pm $0.053 &  16.672$\pm $4.527 &  \textbf{55.693$\pm $3.371} &   0.24$\pm $0.081 \\
		\bottomrule
	\end{tabular}
\end{table}

The LightGBM presents the highest Reduced Precision $(0.38\pm 0.185)$ and the lowest False Alarm $(2.419\pm 0.683)$ in front of the XGBoost. Although it ranks third for Event Recall and average Anticipation Time, behind Naive Bayes and Support Vector Classification, it holds the highest Event F1-score $(0.478\pm 0.189)$.

\subsubsection{Feature Selection with MIG:} 
Each feature was considered with the target variable to calculate the MIG corresponding to them. After normalizing the computed feature importance, features with at least $1\% (0.01)$ importance were selected. The Table \ref{AHE_res_MIG} presents the average results obtained across the 5 folds:

\begin{table}[!h]
	\centering
	\caption{AHE: Average Results After MIG Feature Selection (mean $\pm$ std)}
	\label{AHE_res_MIG}
	\begin{tabular}{llllll}
		\toprule
		{} &             ER &             RP &            aveFA &           aveAT &      EF1-score \\
		Models   &                &                &                  &                 &                \\
		\midrule
		XGBoost  &  0.726$\pm $0.142 &  0.347$\pm $0.167 &      3.176$\pm $0.7 &  48.941$\pm $4.628 &   0.458$\pm $0.19 \\
		LightGBM &  0.703$\pm $0.129 &  \textbf{0.405$\pm $0.177} &    \textbf{2.128$\pm $0.336} &   47.744$\pm $6.39 &  \textbf{0.502$\pm $0.191} \\
		SVC      &   0.68$\pm $0.191 &  0.065$\pm $0.031 &  30.367$\pm $11.069 &  49.455$\pm $6.088 &  0.118$\pm $0.055 \\
		NB       &   \textbf{0.967$\pm $0.03} &  0.114$\pm $0.048 &   22.909$\pm $5.136 &  \textbf{56.798$\pm $1.646} &    0.2$\pm $0.078 \\
		\bottomrule
	\end{tabular}
\end{table}

\newpage

The LightGBM, once again, has the greatest Reduced Precision $(0.405\pm 0.177)$ and the lowest average False Alarms $(2.128\pm 0.336)$. It is in the last position for the average Anticipation Time $(47.744\pm 6.39)$ and in the third position for the Event Recall $(0.703\pm 0.129)$ behind the Naive Bayes and the XGBoost. However, it completely dominates the others with an Event F1-score of $0.502\pm 0.191$.

\subsubsection{Run-time Analysis}
The Table \ref{RunTimeAHE} shows the average running time (in seconds) of the $4$ models for the prediction of Acute Hypotensive Events.

\begin{table}[!h]
	\centering
	\caption{Average Running Time (in seconds) for AHE Prediction (mean $\pm$ std)}
	\label{RunTimeAHE}
	\begin{tabular}{llll}
		\toprule
		{} &       Full Data &  MIG Selection & LightGBM Selection \\
		Models   &                 &                &                    \\
		\midrule
		XGBoost  &  10.382$\pm $4.453 &  8.441$\pm $1.724 &     20.038$\pm $5.507 \\
		LightGBM &    4.24$\pm $1.114 &  1.564$\pm $0.347 &      0.816$\pm $0.123 \\
		SVC      &  11.169$\pm $4.967 &    5.852$\pm $3.1 &      3.204$\pm $1.083 \\
		NB       &   0.263$\pm $0.037 &  0.026$\pm $0.006 &      0.057$\pm $0.013 \\
		\bottomrule
	\end{tabular}
\end{table}

The LightGBM model has then a very low running time, behind the Naive Bayes model.

\subsection{TE Early Prediction}
The Tables \ref{TrainDataTE} and \ref{TestDataTE} show the number of Observation Windows contained in each fold, as well as the number of associated main events, for the Train set and the Test set.

\begin{table}[!h]
	\begin{subtable}[b]{.5\linewidth}
		\centering
		\caption{Train data for TE}
		\label{TrainDataTE}
		\begin{tabular}{llllll}
			\toprule
			{} & Fold 1 & Fold 2 & Fold 3 & Fold 4 & Fold 5 \\
			AHE                 &        &        &        &        &        \\
			\midrule
			OW &   2718 &   2337 &   2497 &   2757 &   2535 \\
			Main Events         &    554 &    519 &    610 &    658 &    435 \\
			\bottomrule
		\end{tabular}
	\end{subtable}
	\hfill
	\begin{subtable}[b]{.5\linewidth}
		\centering
		\caption{Test data for TE}
		\label{TestDataTE}
		\begin{tabular}{llllll}
			\toprule
			{} & Fold 1 & Fold 2 & Fold 3 & Fold 4 & Fold 5 \\
			AHE                 &        &        &        &        &        \\
			\midrule
			OW &    903 &   1277 &    916 &    603 &   1153 \\
			Main Events         &    155 &    380 &     95 &     57 &    285 \\
			\bottomrule
		\end{tabular}
	\end{subtable}
\end{table}

Thus, each fold of the Training set contains at least $2337$ Observation Windows ($2337\ \text{hours}$) with at least $435$ of them related to the Tachycardia Episodes. In addition, each fold of the Test set contains at least $603$ Observation Windows ($603\ \text{hours}$) with at least $57$ of them related to the Tachycardia Episodes.

\subsubsection{Full Dataset:} 
The average results obtained across the 5 folds for the predictive models XGBoost, LightGBM, SVC and Naive Bayes, are presented in the Table \ref{TE_res_full}.

\begin{table}[!h]
	\centering
	\caption{TE: Average Results on the Full Data (mean $\pm$ std)}
	\label{TE_res_full}
	\begin{tabular}{llllll}
		\toprule
		{} &             ER &             RP &           aveFA &           aveAT &      EF1-score \\
		Models   &                &                &                 &                 &                \\
		\midrule
		XGBoost  &  \textbf{0.853$\pm $0.058} &  0.583$\pm $0.099 &   4.843$\pm $1.373 &  \textbf{54.272$\pm $3.466} &  0.691$\pm $0.087 \\
		LightGBM &  \textbf{0.853$\pm $0.045} &  \textbf{0.601$\pm $0.105} &    \textbf{4.323$\pm $1.09} &  53.945$\pm $3.743 &  \textbf{0.703$\pm $0.086} \\
		SVC      &   0.772$\pm $0.11 &  0.229$\pm $0.082 &  20.285$\pm $3.224 &   52.646$\pm $2.21 &  0.342$\pm $0.094 \\
		NB       &  0.633$\pm $0.161 &  0.398$\pm $0.109 &   6.589$\pm $3.604 &  50.676$\pm $4.171 &    0.47$\pm $0.09 \\
		\bottomrule
	\end{tabular}
\end{table}

The LightGBM model shares first place with XGBoost for the Event Recall $(0.853\ \text{in mean})$. But it outperforms all others for Reduced Precision $(0.601\pm 0.105)$, average False Alarms $(4.323\pm 1.09)$ and Event F1-score $(0.703\pm 0.086)$. XGBoost exceeds it in terms of average Anticipation Time ($54.272\pm 3.466$ against $53.945\pm 3.743$).

\subsubsection{Feature Selection with LightGBM:} 
The features with non-zero importance were selected, using the \textit{Split importance} method for the LightGBM feature importance ranking.
The average results obtained across the 5 folds are shown in the Table \ref{TE_res_Light}:

\begin{table}[!h]
	\centering
	\caption{TE: Average Results After LightGBM Feature Selection (mean $\pm$ std)}
	\label{TE_res_Light}
	\begin{tabular}{llllll}
		\toprule
		{} &             ER &             RP &           aveFA &           aveAT &      EF1-score \\
		Models   &                &                &                 &                 &                \\
		\midrule
		XGBoost  &  0.845$\pm $0.053 &  \textbf{0.604$\pm $0.107} &   4.417$\pm $1.217 &  54.303$\pm $2.806 &  \textbf{0.702$\pm $0.088} \\
		LightGBM &  0.851$\pm $0.043 &  0.595$\pm $0.118 &   \textbf{4.232$\pm $1.091} &  53.886$\pm $3.359 &  0.696$\pm $0.095 \\
		SVC      &   0.466$\pm $0.12 &  0.365$\pm $0.198 &  10.014$\pm $6.954 &  50.923$\pm $3.928 &  0.358$\pm $0.087 \\
		NB       &  \textbf{0.895$\pm $0.103} &  0.441$\pm $0.141 &    9.689$\pm $3.12 &   56.32$\pm $2.262 &   0.584$\pm $0.14 \\
		\bottomrule
	\end{tabular}
\end{table}

The LightGBM holds the lowest average False Alarms $(4.232\pm 1.091)$ but only came second for Event Recall $(0.851\pm 0.043)$ behind Naive Bayes, Reduced Precision $(0.595\pm 0.118)$ and Event F1-score $(0.696\pm 0.095)$ behind XGBoost. It is also at the third position for the average Anticipation Time, $(53.886\pm 3.359)$ behind the XGBoost and the Naive Bayes models.

\subsubsection{Feature Selection with MIG:} 
Using the normalized MIG feature importance, features with at least $1\%\ (0.01)$ importance were selected.
The average results obtained across the 5 folds are shown in the Table \ref{TE_res_MIG}:

\begin{table}[!h]
	\centering
	\caption{TE: Average Results After MIG Feature Selection (mean $\pm$ std)}
	\label{TE_res_MIG}
	\begin{tabular}{llllll}
		\toprule
		{} &             ER &             RP &           aveFA &           aveAT &      EF1-score \\
		Models   &                &                &                 &                 &                \\
		\midrule
		XGBoost  &  \textbf{0.853$\pm $0.058} &  0.583$\pm $0.099 &   4.843$\pm $1.373 &  \textbf{54.272$\pm $3.466} &  0.691$\pm $0.087 \\
		LightGBM &  \textbf{0.853$\pm $0.045} &  \textbf{0.601$\pm $0.105} &    \textbf{4.323$\pm $1.09} &  53.945$\pm $3.743 &  \textbf{0.703$\pm $0.086} \\
		SVC      &   0.772$\pm $0.11 &  0.229$\pm $0.082 &  20.285$\pm $3.224 &   52.646$\pm $2.21 &  0.342$\pm $0.094 \\
		NB       &  0.633$\pm $0.161 &  0.398$\pm $0.109 &   6.589$\pm $3.604 &  50.676$\pm $4.171 &    0.47$\pm $0.09 \\
		\bottomrule
	\end{tabular}
\end{table}

The LightGBM model completely dominated the other models with almost all metrics, with notably an Event F1-score of $0.703\pm 0.086$! It ranks second for the average Anticipation Time behind the XGBoost model $(53.945\pm 3.743)$.

\subsubsection{Run-time Analysis}
The Table \ref{RunTimeTE} shows the average running time (in seconds) of the $4$ models for the prediction of Tachycardia Events.

\begin{table}[!h]
	\centering
	\begin{tabular}{llll}
		\toprule
		{} &        Full Data &    MIG Selection & LightGBM Selection \\
		Models   &                  &                  &                    \\
		\midrule
		XGBoost  &   12.641$\pm $1.835 &   12.532$\pm $2.579 &    22.451$\pm $14.734 \\
		LightGBM &     3.173$\pm $0.19 &     2.28$\pm $0.406 &      4.657$\pm $7.151 \\
		SVC      &  49.233$\pm $17.467 &  40.396$\pm $19.571 &     15.941$\pm $8.596 \\
		NB       &     0.402$\pm $0.07 &    0.084$\pm $0.017 &      0.078$\pm $0.036 \\
		\bottomrule
	\end{tabular}
	\caption{Average Running Time (in seconds) for TE Prediction (mean $\pm$ std)}
	\label{RunTimeTE}
\end{table}

Once again, it is observed that the LightGBM model has a very low running time behind the Naive Bayes model.

\section{Discussion}
An early warning approach of potential fatal diseases is an algorithm that provides information about upcoming risks to susceptible people before a Critical Health Episode occurs. This enables action to be taken to reduce possible harm and, in some situations, prevent it from occurring. Cerqueira et al.\cite{ref4} developed a hierarchical approach for the early prediction of both Acute Hypotensive Episodes and Tachycardia Episodes. Since Layered Learning separates a predictive task into two or simpler predictive tasks,
the learning process inside one layer influences the learning process of the following layer. Furthermore, before using the LL approach, a pre-conditional event must be explicitly defined. This procedure is highly domain-dependent. But, end-to-end machine learning techniques are
gaining ground.
The objective of this study was therefore to provide a highly efficient approach for the early prediction of the Acute Hypotensive Episodes and the Tachycardia Episodes. Not only should such an approach have a large average Anticipation Time, but it should also primarily have a very high Event F1-score. This expresses the model's ability to predict Critical Health Episodes earlier with great efficiency, while avoiding false alarms as much as possible.
The proposed AHE and TE early prediction system (MIG-LightGBM) includes the Mutual Information Gain feature selection process and the LightGBM predictive model.

According to the results on the early prediction of AHE, this approach can capture up to $70\%$ of the Acute Hypotensive Events at more than $1$ hour $47$ minutes before their appearance (a gap of $1$ hour had been considered) while maintaining the highest EF1-score of $50\%$ compared to other methods.
When it comes to predicting TE, it can capture up to $85\%$ of the Tachycardia Events at more than $1$ hour $53$ minutes before their appearance while maintaining the highest and remarkable EF1-score of $70\%$.
Although Naive Bayes has a high average Anticipation Time, it has the major drawback of launching several False Alarms as much as it predicts several Critical Health Episodes. The same is true for Support Vector Classification. As for Extreme Gradient Boosting, it does better than the last two, but was overtaken by LightGBM.

The comparison of the performance of the Layered Learning approach (LL) and the MIG-LightGBM method is presented in Table \ref{LLvsLightAHE} for AHE early prediction.

\begin{table}[!h]
	\centering
	\caption{MIG-LightGBM vs LL for AHE Early Prediction}
	\label{LLvsLightAHE}
	\begin{tabular}{llllll}
		\toprule
		{} &                ER &                RP &          aveFA &          aveAT &         EF1-score \\
		\midrule
		MIG-LightGBM &     0.703$\pm $0.129 &     \textbf{0.405$\pm $0.177} &  \textbf{2.128$\pm $0.336} &  \textbf{47.744$\pm $6.39} &     \textbf{0.502$\pm $0.191} \\
		LL           &  \textbf{0.830$\pm $0.054} &  0.205$\pm $0.044 &   3.3$\pm $9.1 &  46.9$\pm $3.3 &  0.328$\pm $0.056 \\
		\bottomrule
	\end{tabular}
\end{table}

The MIG-LightGBM approach completely dominates the LL method with the main efficiency measurement metrics (EF1-score, aveAT, and aveFA). It notably exceeds LL by about $20\%$ on both Event F1-score and Reduced Precision.
\newline
Moreover, LL has an average run-time of $102.3\pm 12.5$ seconds, while MIG-LightGBM has only $1.564\pm 0.347$ seconds.

For TE early prediction, the Table \ref{LLvsLightTE} shows the comparison of the performance of the Layered Learning approach and the MIG-LightGBM method.

\begin{table}[!h]
	\centering
	\caption{MIG-LightGBM vs LL for TE Early Prediction}
	\label{LLvsLightTE}
	\begin{tabular}{llllll}
		\toprule
		{} &                ER &                RP &          aveFA &           aveAT &         EF1-score \\
		\midrule
		MIG-LightGBM &     0.858$\pm $0.054 &     \textbf{0.602$\pm $0.108} &  \textbf{4.285$\pm $1.146} &  \textbf{53.131$\pm $4.182} &     \textbf{0.705$\pm $0.088} \\
		LL           &  \textbf{0.938$\pm $0.027} &  0.136$\pm $0.018 &  35.9$\pm $8.8 &   53.0$\pm $2.2 &  0.237$\pm $0.027 \\
		\bottomrule
	\end{tabular}
\end{table}

As in the case of AHE, the MIG-LightGBM approach completely dominates the LL method with the main efficiency measurement metrics for TE early prediction. It notably exceeds LL by about $50\%$ on both EF1-score and RP.
About the average execution time, LL ($70.9\pm 4.9$) takes more than $68$ seconds longer than MIG-LightGBM ($2.28\pm 0.406$).

The proposed method, MIG-LightGBM, is therefore a highly efficient approach for the early prediction of AHE and TE. Its greatest strength is its ability to identify numerous Critical Health Episodes earlier while avoiding false alarms as much as possible, and more than existing methods. It gives ample time for vital actions to be taken after an impending alarm, and its warnings are highly reliable. Systems that avoid false alarms are highly recommended in healthcare. They avoid alarms fatigue and therefore constitute an excellent improvement in ICU results and the living conditions of nursing staff. Alarm fatigue occurs when busy personnel are subjected to a high volume of regular safety warnings and become insensitive to them. In $2020$, Lewandowska et al.\cite{ref12} investigate on the impact of alarm fatigue on the work of nurses in an intensive care environment. Frequent monitoring alerts may reduce caregivers' capabilities to react when an alarm signals an emergency that demands immediate attention.
This desensitization might result in longer response times or the ignoring of vital alerts, causes of deaths in ICUs.

However, MIG-LightGBM has a main limitation. By trying to avoid false alarms as much as possible, it can lose sight of a few Critical Health Episodes.
This study focused much more on improving the model's ability to avoid false alarms while predicting CHEs early and efficiently. To perfect the proposed warning system, it would also be good to significantly increase the model's ability not to miss the CHEs while maintaining the current strength.
\section{Conclusion}

This study focused on building a highly effective early warning system for the Critical Health Episodes such as Acute Hypotensive and Tachycardia. This system is able to predict them earlier with great efficacy, while avoiding false alarms as much as possible. With a high Reduced Precision, a large average Anticipation Time and a very strong Event F1-score on the MIMIC II dataset, the proposed method, MIG-LightGBM, dominates the existing approaches. Its performance is therefore reliable. In practical situations, it can be applied on patients to make a reliable early prediction on Critical Health Episodes, having knowledge of at least their latest one-hour observations. This system also has the benefit of being able to avoid alarms fatigue.
But because of its Event Recall, which is not extremly high, it can sometimes omit some Critical Health Episodes. A direct future work could therefore focus on strengthening the system's ability to identify CHEs without losing its current strength.

\section{Data availability}
The original data was provided by Cerqueira et al.\cite{ref4} at \url{https://github.com/vcerqueira/layered_learning_time_series/tree/master/data_sample}. Patient records, obtained after the feature engineering process we performed, can be accessed at \url{https://drive.google.com/drive/folders/1X14qiqlIYU_YNcbSMnKIcXttb2_FReMK?usp=sharing}.

%
%
%
%

\end{document}